# CoRe: An Automated Pipeline for The Prediction of Liver Resection Complexity from Preoperative CT Scans

Omar Ali[1-4], Alexandre Bône[1], Caterina Accardo[5], Omar Belkouchi[5], Marc-Michel Rohé[1], Eric Vibert[2,4,5], Irene Vignon-Clementel[3]

[1] Guerbet Research, Villepinte, France
[2] Inserm U1193, Villejuif, France
[3] Research Center Inria Saclay Ile-de-France, Palaiseau, France
[4] Paris – Saclay University, Gif-sur-Yvette, France
[5] Paul Brousse Hospital – APHP, Villejuif, France
`omar.ali@guerbet.com`

**Abstract.** Surgical resections are the most prevalent curative treatment for primary liver cancer. Tumors located in critical positions are known to complexify liver resections (LR). While experienced surgeons in specialized medical centers may have the necessary expertise to accurately anticipate LR complexity, and prepare accordingly, an objective method able to reproduce this behavior would have the potential to improve the standard routine of care, and avoid intra- and postoperative complications. In this article, we propose CoRe, an automated medical image processing pipeline for the prediction of postoperative LR complexity from preoperative CT scans, using imaging biomarkers. The CoRe pipeline first segments the liver, lesions, and vessels with two deep learning networks. The liver vasculature is then pruned based on a topological criterion to define the hepatic central zone (HCZ), a convex volume circumscribing the major liver vessels, from which a new imaging biomarker, $B_{HCZ}$ is derived. Additional biomarkers are extracted and leveraged to train and evaluate a LR complexity prediction model. An ablation study shows the HCZ-based biomarker as the central feature in predicting LR complexity. The best predictive model reaches an accuracy, F1, and AUC of 77.3, 75.4, and 84.1% respectively.

**Keywords:** Vasculature Analysis, Imaging Biomarkers, Surgical Resection Complexity Prediction

## 1    Introduction

Liver cancer is a prominent contributor to cancer mortality worldwide, ranking second in the most common causes of cancer-related deaths [1]. In the management of the early stages of liver cancer, liver resection (LR) is the most prevalent type of treatment [1]. However, with considerable variations in the technicalities of different types of LRs, a preoperative assessment of resection complexity is necessary to minimize the intra- and postoperative risks [2].

In the medical literature, LR complexity can be evaluated at the pre-, intra-, and postoperative stages of the surgical operation [2-4]. While the preoperative complexity is based on the evaluation of the different types of LRs [3], the intra- and postoperative LR complexities rely on the surgical maneuvers performed during the operation [2], [4] and on the expertise of the liver surgeon [2], [3]. However, across all these definitions, medical expertise and user interactions are required.

Furthermore, centrally located liver tumors require more technically challenging surgical resections due to their proximity to the major hepatic vessels (portal and hepatic veins) [5], often necessitating vasculature reconstruction, which according to [6] further increases the complexity of the resection. Therefore, a detailed knowledge of the morphology and structure of the hepatic vasculature is necessary to assess the position of the tumors with respect to the major hepatic vessels. The methods proposed in [7-8] present a semi-automatic approach for the geometrical and structural analysis of the liver vessels in the context of preoperative liver surgery planning.

In this article, we introduce CoRe, the first automatic, quantitative, and interpretable pipeline, for the prediction of the postoperative LR complexity from preoperative portal phase CT scans, using imaging-based biomarkers. First, segmentations of the liver, lesions, and hepatic vessels are generated with a state-of-the-art UNet deep network, benefitting from the literature in [9-14]. Second, we propose a liver vasculature pruning algorithm to define the "hepatic central zone" (HCZ), a convex volume circumscribing the major liver vessels, from which a new imaging biomarker, $B_{HCZ}$ is derived. Third, additional quantitative biomarkers are leveraged from the generated segmentations to train and evaluate a LR complexity prediction model.

## 2 Methods

### 2.1 Liver, Lesion, and Vessel Segmentation

The details of the CoRe pipeline are depicted in Figure 1. The first step is the segmentation of the liver, lesions, and vessels from the portal phase of preoperative CT scans. Two 3D convolutional neural networks (CNN) are trained to segment the liver and liver lesions on one hand, and the portal and hepatic vessels on the other hand. The segmentations are then combined by appropriately labeling their union (Fig. 1A).

We leverage the state-of-the-art nnUNet's [14] default 3D full resolution framework to train both segmentation models. These models follow a UNet-like architecture [15], and are trained with the dice and cross-entropy losses using stochastic gradient descent with Nesterov momentum, and a geometrically decaying learning rate. The segmentation models are denoted LivLes3D and HepVess3D respectively.

### 2.2 Topological Analysis of the Liver Vasculature

The predicted binary vessel segmentations are pruned to keep the major vessels only, and define the HCZ (Fig. 1B). The successive steps are listed below.

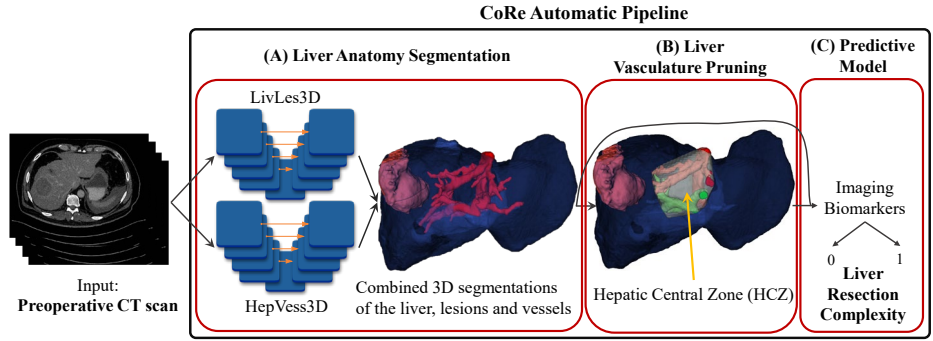

**Fig. 1.** Details of the CoRe pipeline. (A): Liver anatomy segmentation results with liver (blue), lesion (orange) and vessels (red). (B): Liver vasculature pruning algorithm (green: portal veins, red: hepatic veins), and HCZ generation (white volume around the vessels inside the liver). (C): Predictive model of LR complexity from imaging biomarkers extracted from (A) and (B), where 0 and 1 refer to not complex and complex cases respectively.

1. **Skeletonization**. The segmentation mask is skeletonized with Scikit-Image 0.20.0 using [16]. Local vessel radii are associated to each skeleton element after computing the distance transform of the vessel masks with Scipy 1.8.0.
2. **Graph construction**. The skeleton is converted into a graph representation $G = (V, E)$ with vertices $V = \{v_1, \ldots, v_N\}$ and undirected edges $E = \{(v, w) \mid v, w \in V, v \neq w\}$, considering two voxels as neighbors if their corresponding indexes differ by a maximum value of 1 in each direction.
3. **Branch decomposition**. $G$ is decomposed into edge-disjoint subgraphs $B_1, \ldots, B_M$ called branches, such that each edge $(v, w) \in E$ belongs to exactly one branch, using Skan 0.11.0 [17]. The length $len(B)$ and mean radius $rad(B)$ of each branch $B$ are computed, either by summation or averaging the edge length and radius respectively.
4. **Vascular entry identification.** Anatomically, the liver vasculature is composed of two distinct vessel trees, the portal and hepatic trees. To identify the corresponding liver entry points, the vessel mask is successively eroded, and the centroids of the two most persistent connected components are projected onto the skeleton. The closest vertices $v_p$ and $v_h$ with degree 1 are then identified.
5. **Tree pruning**. See Algorithm 1 below. Two tree hierarchies are extracted from $G$ using $v_p$ and $v_h$ as respective seeding vertices, retaining only the major connected branches. The branch lengths and diameters are exploited to identify significant branch bifurcation levels $bif(B)$ in a recursive tree building approach. The vessels extending beyond the third major vascular bifurcation are considered irrelevant for surgical planning and are pruned.
6. **Morphological reconstruction**. The retained skeletons are inflated by a dilation factor equal to the local diameters. Numerical round-up and mask intersection with the original segmentation allow to reconstruct the local vessel geometry.
7. **Hepatic central zone definition**. The convex hull of the retained vessels defines the hepatic central zone (HCZ).

**Algorithm 1.** Vessel tree pruning algorithm.

---

Inputs:
- **graph** $G = (V, E)$, **branches** $(B_m)_m$, **root** $v_r \in V$;
- current **bifurcation level** $bif(B)$, **length** $len(B)$ and **radius** $rad(B)$;
- archive of **visited subgraph** $G_{visited}$.

Hyperparameters:
- maximum vessel tree **bifurcation level** $bif_{max} = 2$;
- maximum vessel branch **reduction factor** $R_{max} = 0.2$.

---

Let $w_1, \ldots, w_d \in V$ be the neighbor vertices of $v_r$, where $d \in \mathbb{N}$ is the degree of $v_r$.
**If** $bif(B)$, $len(B)$ **and** $rad(B)$ are undefined:   # initial call, d=1
    **Identify** the branch $B^{w_1}$ such as $(v_r, w_1) \in B$.
    **Initialize** $len(B) \leftarrow len(B^{w_1})$ and $rad(B) \leftarrow rad(B^{w_1})$.
    **Initialize** $bif(B) \leftarrow 0$ and $G_{visited}$ with $v_r$.
**If** $bif(B) = bif_{max}$ **or** all $w_1, \ldots, w_d$ belong to $G_{visited}$:
    **Return** $G_{visited}$.
**Else if** there is a single neighbor vertex $w$ that does not belong to $G_{visited}$:
    **Return** $G_{visited} \leftarrow$ recursive call using $w$ as root.
**Else**, **for each** neighbor vertex $w$ that does not belong to $G_{visited}$:
    **Identify** the branch $B^w$ such as $(v_r, w) \in B$.
    **If** $len(B^w) < R_{max} \cdot len(B)$ **or** $rad(B^w) < R_{max} \cdot rad(B)$:   # noise
        **Update** $G_{visited} \leftarrow$ recursive call using $w$ as root.
    **Else**:   # relevant branch
        **Update** $G_{visited} \leftarrow$ recursive call using $w$, $bif(B) + 1$, $len(B^w)$, $rad(B^w)$.
    **Return** $G_{visited}$.

---

Output: **pruned vessel subgraph** $G_{visited}$.

---

This vasculature analysis approach builds on [8] to propose an entirely automated imaging pipeline where the segmentation noise is handled during the tree construction step, via filtering heuristics based on vessel graph properties. The tree construction procedure is detailed in Algorithm 1.

### 2.3   Quantitative Imaging Biomarkers for LR Complexity Prediction

The liver volume $V_{Liv}$, the number of lesions $N_{Les}$, and the volume of lesion $V_{Les}$ are the imaging biomarkers directly extracted from the raw segmentations. We also define a new imaging biomarker $B_{HCZ}$ based on the proposed HCZ, as the lesions' relative occupancy volume $V$ inside the HCZ if it is nonzero, and the negative of the minimal distance $d$ from the lesion to the HCZ. $B_{HCZ}$ is defined in equation 1 below.

$$B_{HCZ} = \begin{cases} \dfrac{V_{Les \cap HCZ}}{V_{HCZ}} & if \quad Les \cap HCZ \neq \emptyset \\ -\dfrac{d(Les, HCZ)}{diameter(HCZ)} & otherwise \end{cases} \quad (1)$$

The liver volume is used for its importance in withstanding the bodily functions post-LR. The volume and the number of lesions are considered, since larger or multiple lesions are generally associated with complex LRs. Lesions in the vicinity of the HCZ are assumed to be located in critical positions given their proximity to the liver's major vessels. Accordingly, the proposed biomarker $B_{HCZ}$ is utilized as a quantitative indicator of the position of the liver lesions with respect to the HCZ.

Lastly, with scikit-learn v0.23.2, the default configuration of the logistic regression is chosen as the binary classifier to train a cross-entropy loss with an L2 regularization using the biomarkers above, in order to predict LR complexity (Fig. 1C).

## 3 Experiments

### 3.1 Datasets and Preprocessing

IRCAD [18] and LiTS [19], are public datasets of abdominal CT images with liver, lesion, and hepatic vessel (available with IRCAD only) reference segmentations. LiTS also provides a benchmark validation set ($LiTS_{VS}$) with 70 unannotated scans for the evaluation of the liver and lesion segmentations. Moreover, an internal dataset ($d_{seg} = 65$) with manually segmented vessel annotations is used as a complementary training set for HepVess3D.

IRCAD, LiTS, and $d_{seg}$ are preprocessed using the nnUNet's default preprocessing pipeline which includes a contrast clipping, a z-score normalization, and a resampling of the images to their median spacing [14].

An internal dataset ($d_{internal}$) of 128 patients who underwent LR for liver cancer was created between 2012-2020 as a separate internal dataset to assess the proposed pipeline. Labeling LR complexity is a regular clinical practice in our medical institution, and is assigned postoperatively on a 1-10 scale by the surgeon who performed the LR. Following [2], and accounting for the surgical triage application, two groups identified to be the 'not complex' and 'complex' groups were considered. Accordingly, the scores are binarized at the 5 threshold where the values 0 and 1 refer to the 'not complex' and 'complex' cases respectively. Overall, the dataset is balanced with 63 complex and 65 not complex LRs.

### 3.2 Training, Evaluation, and Inference

LivLes3D is trained for 1000 epochs on the combination of IRCAD and LiTS with an initial learning rate $lr_{init} = 0.01$, and evaluated on $LiTS_{VS}$ through their online platform. HepVess3D is also trained for 1000 epochs with $lr_{init}$ on the combination of $d_{seg}$ and IRCAD, using 5-fold cross validation only on IRCAD's hepatic vessels. The performance of both models is evaluated using the dice metric.

In inference, LivLes3D and HepVess3D are employed on $d_{internal}$ to generate liver, lesions, and hepatic vessel segmentations from which the imaging biomarkers defined in 2.3 are automatically extracted. Lastly, the LR complexity predictive model is trained and evaluated on $d_{internal}$ using the leave-one-out method.

Table 1. Results of the ablation study in %. In bold are the best results.

| $B_{HCZ}$ | $N_{Les}$ | $V_{Les}$ | $V_{Liv}$ | Accuracy | F1 | AUC |
|---|---|---|---|---|---|---|
| ✓ | ✓ | ✓ | ✓ | **73.4 (3.9)** | **71.2 (4.8)** | **83.5 (3.5)** |
| ✗ | ✓ | ✓ | ✓ | 70.3 (4.0) | 67.8 (5.0) | 78.2 (4.0) |
| ✓ | ✗ | ✓ | ✓ | 75.8 (3.8) | 73.2 (4.8) | 79.0 (4.1) |
| ✓ | ✓ | ✗ | ✓ | 72.6 (4.0) | 71.4 (4.7) | 82.0 (3.7) |
| ✓ | ✓ | ✓ | ✗ | **77.3 (3.5)** | **75.4 (4.0)** | **84.1 (3.6)** |
| ✗ | ✓ | ✓ | ✗ | 71.8 (4.0) | 70.2 (4.7) | 77.8 (4.1) |
| ✓ | ✗ | ✓ | ✗ | 73.4 (3.9) | 70.4 (5.0) | 78.8 (4.1) |
| ✓ | ✓ | ✗ | ✗ | **77.8 (3.7)** | **78.1 (4.2)** | **81.6 (3.9)** |
| ✗ | ✓ | ✗ | ✗ | 66.4 (4.1) | 60.0 (5.6) | 65.2 (5.0) |
| ✓ | ✗ | ✗ | ✗ | **66.4 (4.2)** | **65.3 (5.0)** | **72.8 (4.4)** |

An ablation study is carried out to determine the best set of biomarkers for predicting LR complexity. An iterative backward elimination is performed on the least impactful feature, starting from the baseline configuration with the four predefined biomarkers as inputs. The performance of the models is evaluated using the accuracy, F1, and AUC metrics.

## 4 Results

### 4.1 Quantitative Results

**Segmentation Performance**. The mean dice scores obtained on the different benchmark test sets (see 3.1) are in-line with the models in the literature [9-14], achieving 96.0%, and 71.6% for the liver, and lesion segmentations on $LiTS_{VS}$ (standard deviation not reported on LiTS online platform), and 79.1 ± 4.0% for vessel segmentations on IRCAD respectively. With the unavailability of reference annotations for $d_{internal}$, the segmentations generated by LivLes3D and HepVess3D were exhaustively verified by expert liver surgeons, confirming satisfactory segmentation results on this dataset.

**Liver Resection Complexity Prediction.** The results of the ablation study (see 3.2) are reported in Table 1, and Fig. 2. The baseline configuration with $B_{HCZ}$, $N_{Les}$, $V_{Les}$, and $V_{Liv}$ reaches an accuracy, F1, and AUC of 73.4, 71.2, and 83.5% respectively. With continuous feature ablation, the best configuration for predicting LR complexity combines the HCZ based biomarker $B_{HCZ}$ with $N_{Les}$, and $V_{Les}$, achieving an accuracy, F1, and AUC of 77.3, 75.4, and 84.1% respectively. Additionally, the first feature that can be disregarded for having the least impact on the prediction of LR complexity is $V_{Liv}$ followed $V_{Les}$ and $N_{Les}$. The proposed biomarker $B_{HCZ}$ manifests as the central feature for predicting LR complexity, achieving the best scores in single feature evaluation with an accuracy, F1, and AUC of 66.4, 65.3 and 72.8% respectively.

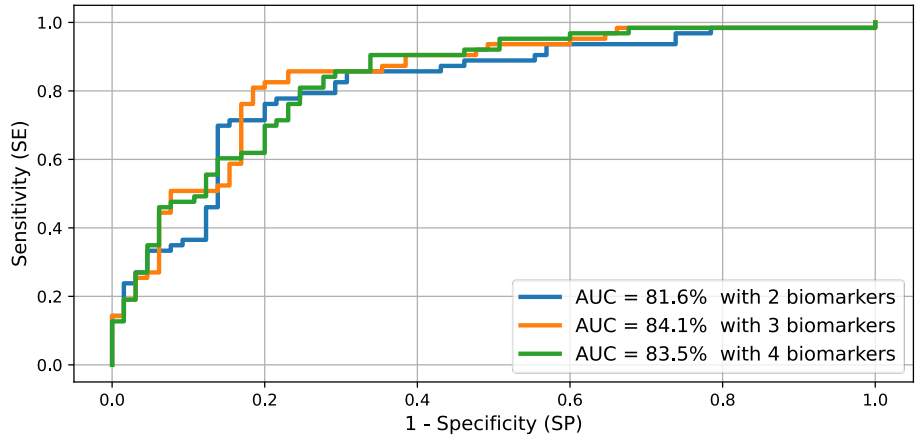

**Fig. 2.** ROC curves from the ablation study showing the best three LR complexity predictive models with multiple input biomarkers. In blue, the biomarkers are: $F_{HCZ}$, and $N_{Les}$. In orange, the biomarkers are: $F_{HCZ}$, $N_{Les}$, $V_{Les}$. In green, the features are: $F_{HCZ}$, $N_{Les}$, $V_{Les}$ and $V_{Liv}$.

### 4.2 Qualitative Results

**Liver Anatomy Segmentation, Vessel Pruning, and HCZ Generation**. Fig. 3 shows the segmentation results of the liver, lesions, and vessels for four distinct cases in $d_{internal}$, along with the results of the vessel pruning and HCZ generation. The most important biomarker for the prediction of LR complexity in cases (A), (B), and (C) is the imaging biomarker $B_{HCZ}$. Cases (B) and (C) are labeled as complex LRs, and are correctly predicted with $B_{HCZ}$ values at 98 and 40 respectively, reflecting the lesions' occupancy volumes in the HCZ. In case (A), correctly predicted as not complex, $B_{HCZ} = -40.8$ reflects the negative minimal distance of the lesion with respect to the HCZ. Lastly, case (D) shows a mispredicted borderline case, where $B_{HCZ} = -55.3$, yet the surgery was labeled complex after binarization (initial label 6/10).

## 5 Discussion and Conclusion

In this article, we presented CoRe, a fully automatic segmentation and postprocessing pipeline to predict postoperative LR complexity from preoperative CT scans. The predictive biomarkers used in this study solely rely on morphological and topological aspects extracted from the segmentations. A new imaging biomarker was introduced to capture the position of the lesions with respect to the highly vascularized HCZ, where surgical LRs are assumed to be complex. It is compared to the classical imaging biomarkers used in surgical planning identified to be the liver and lesions volumes and the number of lesions. Overall, in line with the medical practice, the results confirm the importance of analyzing the position of the liver lesions with respect to the major liver vessels, and the proposed HCZ-based biomarker proved to be central in predicting LR complexity. Moreover, the results clearly show that the proposed bio-marker outperforms the traditional imaging biomarkers in the evaluation of LR complexity.

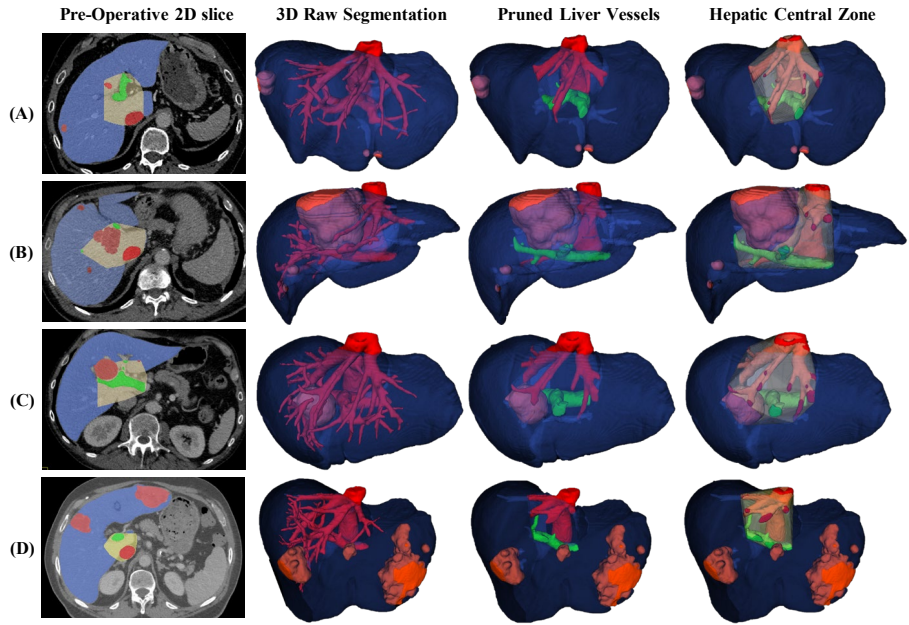

**Fig. 3.** Four cases at different steps in the CoRe pipeline with varying lesion positions with respect to the HCZ. Same color coding as Fig. 1.

A limitation of CoRe is its feed-forward architecture, which can favor the accumulation of errors across the different algorithmic blocks. Yet, it is mitigated by the pipeline's interpretability and explainability: with CoRe, visual verifications are possible at every intermediate step.

In future work, further investigation of the biomarker types (i.e., radiomics, patient demographics), the categorization of LR complexity scores, and the extraction of additional imaging biomarkers such as the distance between the main lesion and its proximal lesions will be considered.

## Acknowledgements

This work was supported by Guerbet and the Ile-de-France Region.